\documentclass[aps,prb,twocolumn,groupedaddress,showpacs]{revtex4-1}
\newcommand{\bec}[1]{\mbox{\boldmath $ #1$}}
\usepackage{graphicx}
\begin{document}
\title{Experimental study of temperature fluctuations in
forced stably stratified turbulent flows}
\author{A. Eidelman}
\email{eidel@bgu.ac.il}
\author{T. Elperin}
\email{elperin@bgu.ac.il}
\homepage{http://www.bgu.ac.il/me/staff/tov}
\author{I. Gluzman}
\email{gluzmany@post.bgu.ac.il}
\author{N. Kleeorin}
\email{nat@bgu.ac.il}
\author{I. Rogachevskii}
\email{gary@bgu.ac.il} \homepage{http://www.bgu.ac.il/~gary}
\affiliation{The Pearlstone Center for Aeronautical Engineering
Studies, Department of Mechanical Engineering, Ben-Gurion University
of the Negev, P.O.Box 653, Beer-Sheva 84105,
Israel}
\date{\today}
\begin{abstract}
We study experimentally temperature fluctuations
in stably stratified forced turbulence in air
flow. In the experiments with an imposed vertical
temperature gradient, the turbulence is produced
by two oscillating grids located nearby the side
walls of the chamber.  Particle Image Velocimetry
is used to determine the turbulent and mean
velocity fields, and a specially designed
temperature probe with sensitive thermocouples is
employed to measure the temperature field. We
found that the ratio $\big[(\ell_x \nabla_x
\overline{T})^2 + (\ell_y \nabla_y
\overline{T})^2 + (\ell_z \nabla_z
\overline{T})^2\big] / \langle \theta^2 \rangle$
is nearly constant, is independent of the
frequency of the grid oscillations and has the
same magnitude for both, stably and unstably
stratified turbulent flows, where $\ell_i$ are
the integral scales of turbulence along $x, y, z$
directions, $\overline{T}$ and $\theta$ are the
mean and turbulent fluctuations components of the
fluid temperature. We demonstrated that for large
frequencies of the grid oscillations, the
temperature field can be considered as a passive
scalar, while for smaller frequencies the
temperature field behaves as an active field. The
theoretical predictions based on the budget
equations for turbulent kinetic energy, turbulent
potential energy ($\propto \langle \theta^2
\rangle)$ and turbulent heat flux, are in a good
agreement with the experimental results. Detailed
comparison with experimental results obtained
previously in unstably stratified forced
turbulence is performed.
\end{abstract}


\maketitle

\section{Introduction}

In the last two decades, the theory of stably
stratified flows undergoes essential revision.
Since the classical papers by Kolmogorov, Obukhov
and Heisenberg, practically used turbulence
closures for the neutrally and stably stratified
flows describe the energetics of turbulence using
the budget equation for the turbulent kinetic
energy (TKE) in combination with the Kolmogorov's
hypotheses for the dissipation rate $\sim
E_K^{3/2} / \ell$ , the eddy viscosity (or the
eddy conductivity and eddy diffusivity)
proportional to $\sim E_K^{1/2} \ell$  (see Ref.
\cite{MY71}), where $E_K$ is the turbulent
kinetic energy and $\ell$ is the mixing length
scale. The straightforward application of this
approach for stably stratified shear flows  led
to the turbulence cut off at Richardson numbers,
Ri $= (N/\overline{S})^2$, exceeding some
critical value, assumed to be close to the
conventional linear instability threshold from
1/4 to 1 (see, e.g., Refs. \cite{CHA61,MIL86}).
The latter assertion, however, contradicts to
experimental evidence and experience from
numerical modelling (see, e.g., Refs.
\cite{SEKR09,BAN02,PAR02,MO02,LA04}). Here $N$ is
the Brunt-V\"{a}is\"{a}l\"{a} frequency
determined by the vertical derivative of the mean
temperature (or the mean potential temperature),
i.e., $N^2=\beta \nabla_z \overline{T}$,
$\beta=g/T_\ast$ is the buoyancy parameter, $g =
9.81$ m/s is the acceleration due to gravity,
$T_\ast$ is a reference value of the absolute
mean temperature $\overline{T}$ and
$\overline{S}$ is the mean shear.

Over decades in meteorology, this difficulty was
overcome heuristically – importing empirical
Ri-dependent coefficients in the expressions for
the eddy viscosity and eddy conductivity.
Recently an insight into this long-standing
problem (since Richardson\cite{RI20}) has been
gained through more rigorous analysis of the
turbulent energetics involving additional budget
equation – for the turbulent potential energy
(TPE) conceptually similar to the Lorenz's
available potential energy\cite{LO67}, and
accounting for the energy exchange between TKE
and TPE (see, Refs.
\cite{EKR05,ZKR07,ZKR08,ZKR09,ZKR12}). This
analysis uses the conservation law for the total
turbulent energy (TTE = TKE+TPE), the budget
equation for turbulent heat flux and opens new
prospects toward developing consistent and
practically useful turbulent closures based on a
minimal set of equations. This approach results
in the asymptotically linear Ri-dependence of the
turbulent Prandtl number and removes the
puzzling, almost the-century old problem of the
unrealistic turbulence cut off (implying the
existence of a critical Richardson number).

In contrast to meteorology the energy exchange
between TKE and TPE was discussed long ago in the
context of the oceanic stably stratified
turbulence\cite{OT87} (see also Refs.
\cite{T77,H86,CM93,SG95,KV00,LCU02,Z02,JSG03,HH04,U05,RH05,MSZ07,LPR08,LR08}).
Detailed discussions of the state of the art in
the turbulence closure problem for stably
stratified flows can be found in Refs.
\cite{ZKR07,ZKR08,ZKR09,CAN09,CAN08}.

The above discussed new ideas should be
comprehensively investigated and validated using
laboratory experiments and numerical simulations
in different set-ups. The goal of this study is
to conduct a comprehensive experimental
investigation of heat transport in temperature
stratified forced turbulence.  In the experiments
turbulence is produced by the two oscillating
grids located nearby the side walls of the
chamber. We use Particle Image Velocimetry to
determine the velocity field, and a specially
designed temperature probe with sensitive
thermocouples is employed to measure the
temperature field. Similar experimental set-up
and data processing procedure were used
previously in the experimental study of different
aspects of turbulent convection (see Refs.
\cite{BEKR11,EEKR06}) and in Refs.
\cite{BEE04,EE04,EEKR06C,EEKR06A,EKR10} for
investigating a phenomenon of turbulent thermal
diffusion. \cite{EKR96,EKR97} Comprehensive
investigation of turbulent structures, mean
temperature distributions, velocity and
temperature fluctuations can elucidate a
complicated physics related to particle
clustering and formation of large-scale
inhomogeneities in particle spatial distributions
in stably stratified turbulent flows.

In the present study we perform a detailed
comparison with experimental results obtained
recently in unstably stratified forced
turbulence\cite{BEKR11}, whereby transition
phenomena caused by the external forcing (i.e.,
transition from Rayleigh-B\'{e}nard convection
with the large-scale circulation (LSC) to the
limiting regime of unstably stratified turbulent
flow without LSC where the temperature field
behaves like a passive scalar) have been studied.
In particular, when the frequency of the grid
oscillations is larger than a certain value, the
large-scale circulation in turbulent convection
is destroyed, and the destruction of the LSC is
accompanied by a strong change of the mean
temperature distribution. However, in all regimes
of the unstably stratified turbulent flow the
ratio $\big[(\ell_x \nabla_x \overline{T})^2 +
(\ell_y \nabla_y \overline{T})^2 + (\ell_z
\nabla_z \overline{T})^2\big] / \langle \theta^2
\rangle$ varies slightly (even in the range of
parameters whereby the behaviour of the
temperature field is different from that of the
passive scalar).\cite{BEKR11}

This paper is organized as follows. The
theoretical predictions are given in Section II.
Section III describes the experimental set-up and
instrumentation. The results of laboratory study
of the stably stratified turbulent flow and
comparison with the theoretical predictions are
described in Section IV. Finally, conclusions are
drawn in Section V.

\section{Theoretical predictions}

In our theoretical analysis we use three budget
equations for the turbulent kinetic energy
$E_k=\langle {\bf u}^2 \rangle/2$,  for the
temperature fluctuations $E_\theta=\langle
\theta^2 \rangle/2$ and for the turbulent heat
flux $F_i = \langle u_i \theta \rangle$:
\begin{eqnarray}
{DE_k \over Dt} + {\rm div} \, {\bf \Phi}_{k} &=&
-\langle u_i \, u_j \rangle \, \nabla_j
\overline{U}_i + \langle {\bf u} {\bf \cdot} {\bf
f}_f \rangle +\beta \, F_z  - \varepsilon_k ,
\nonumber\\
 \label{FA2}\\
{D E_\theta \over Dt} + {\rm div} \, {\bf
\Phi}_\theta &=& - ({\bf F} {\bf \cdot}
\bec{\nabla}) \overline{T} - \varepsilon_\theta ,
 \label{A1}\\
{D F_i \over Dt} + {\nabla}_j \, {\bf
\Phi}_{ij}^{(F)} &=& \beta_i \, \langle \theta^2
\rangle - {1\over \rho} \, \langle \theta \,
\nabla_i p \rangle - \langle u_i u_j \rangle \,
{\nabla}_j \, \overline{T}
\nonumber\\
&&- ({\bf F} {\bf \cdot} \bec{\nabla})
\overline{U}_i - \varepsilon_i^{(F)},
 \label{A11}
\end{eqnarray}
(see, e.g., Refs. \cite{KF84,CCH02,ZKR07,ZKR08}),
where $D / Dt =
\partial /\partial t + \overline{\bf U} {\bf \cdot}
\bec{\nabla}$, $\, {\bf u}$ are the fluctuations
of the fluid velocity, $\theta$ are the
temperature fluctuations, $\overline{\bf U}$ is
the mean velocity, $\overline{T}$ is the mean
temperature, $p$ are the pressure fluctuations,
$\beta_i = \beta \, e_i$, $\, {\bf e}$ is the
vertical unit vector and $\rho$ is the fluid
density. The terms ${\bf \Phi}_{k}$, ${\bf
\Phi}_\theta$ and ${\bf \Phi}_{ij}^{(F)}$ include
the third-order moments. In particular, ${\bf
\Phi}_{k} = \rho^{-1} \langle {\bf u} \, p\rangle
+ (1/2) \langle {\bf u} \, {\bf u}^2 \rangle$
determines the flux of $E_K$, $\, {\bf
\Phi}_\theta = \langle {\bf u} \, \theta^2
\rangle/2$ determines the flux of $E_\theta$ and
$\, {\bf \Phi}_{ij}^{(F)} = \langle u_i u_j
\theta\rangle + \delta_{ij} \rho^{-1} \, \langle
\theta \, p \rangle / 2$ determines the flux of
${\bf F}$. The term $\langle {\bf u} {\bf \cdot}
{\bf f}_f \rangle$ in Eq.~(\ref{FA2}) determines
the production rate of turbulence caused by the
grid oscillations and $\varepsilon_k$ is the
dissipation rate of the turbulent kinetic energy.
The term $\varepsilon_\theta$ in Eq.~(\ref{A1})
determines the dissipation rate of $E_\theta$,
while the term $\varepsilon_i^{(F)}$ in
Eq.~(\ref{A11}) is the dissipation rate of the
turbulent heat flux.

By means of Eq.~(\ref{A1}) we arrive at the
evolutionary equation for the turbulent potential
energy $E_p= (\beta^2 / N^2) \, E_\theta$:
\begin{eqnarray}
{D E_p \over Dt} + {\rm div} \, {\bf \Phi}_p =P_p
- \beta F_z - \varepsilon_p ,
 \label{PA1}
\end{eqnarray}
(see, e.g., Refs. \cite{ZKR07,ZKR08}), where
$N^2= \beta \, \nabla_z \overline{T}$, $\, {\bf
\Phi}_p= (\beta^2 /N^2) \, {\bf \Phi}_\theta$,
$\, P_p= - (\beta^2 /N^2) \, ({\bf F}_h {\bf
\cdot} \bec{\nabla}) \overline{T}$ is the source
(or sink) of the turbulent potential energy
caused by the horizontal turbulent heat flux
${\bf F}_h=\langle {\bf u}_h \, \theta \rangle$,
$\, {\bf u}_h$ is the horizontal component of the
velocity fluctuations and $\varepsilon_p=
(\beta^2 /N^2) \, \varepsilon_\theta$. The
buoyancy term, $\beta \, F_z$, appears in
Eqs.~(\ref{FA2}) and~(\ref{PA1}) with opposite
signs and describes the energy exchange between
the turbulent kinetic energy and the turbulent
potential energy. These two terms cancel in the
budget equation for the total turbulent energy,
$E=E_k+E_p$:
\begin{eqnarray}
{D E \over Dt} + {\rm div} \, {\bf \Phi} &=& P_p
-\langle u_i \, u_j \rangle \, \nabla_j
\overline{U}_i + \langle {\bf u} {\bf \cdot} {\bf
f}_f \rangle - \varepsilon ,
 \label{PA2}
\end{eqnarray}
where ${\bf \Phi}={\bf \Phi}_k + {\bf \Phi}_p$
and $\varepsilon= \varepsilon_k + \varepsilon_p$.
The concept of the total turbulent energy is very
useful in analysis of stratified turbulent flows.
In particular, it allows to elucidate  the
physical mechanism for the existence of the shear
produced turbulence for arbitrary values of the
Richardson number, and abolish the paradigm of
the critical Richardson number in the stably
stratified atmospheric turbulence (see Refs.
\cite{ZKR07,ZKR08}).

Now we use the budget equation~(\ref{A11}) for
the turbulent heat flux $F_i = \langle u_i \theta
\rangle$. According to the estimate made in Ref.
\cite{ZKR07}, $\beta_i \, \langle \theta^2
\rangle - \rho^{-1} \, \langle \theta \, \nabla_i
p \rangle \approx C_\theta \beta_i \, \langle
\theta^2 \rangle$, where $C_\theta$ is an
empirical constant. In a steady-state case
Eq.~(\ref{A11}) yields the components of the
turbulent heat flux $F_x = - D^T_x \, \nabla_x
\overline{T}$, $F_y = - D^T_y \, \nabla_y
\overline{T}$ and $F_z = - D^T_z \, \nabla_z
\overline{T} + C_\theta\, C_F \, \tau_0 \,\beta
\, \langle \theta^2 \rangle$, where $D^T_i = C_F
\, \tau_0 \, \langle u_i^2 \rangle$ with $i=x, y,
z$ are the turbulent temperature diffusion
coefficients in $x, y$, and $z$ directions and
$C_F$ is an empirical constant. Here we have
taken into account that the dissipation rate of
the turbulent heat flux is $\varepsilon_i^{(F)} =
F_i / C_F \tau_0$, the diagonal components of the
tensor $\langle u_i u_j \rangle$ are much larger
than the off-diagonal components, and $\tau_x
\approx \tau_y \approx \tau_z = \tau_0$.

On the other hand, in a steady-state case
Eq.~(\ref{A1}) yields
\begin{eqnarray}
\langle \theta^2 \rangle=- 2 \, \tau_0 \,({\bf F}
{\bf \cdot} \bec{\nabla}) \overline{T},
\label{AB1}
\end{eqnarray}
where we have taken into account that the
dissipation rate of $E_\theta=\langle \theta^2
\rangle /2$ is $\varepsilon_\theta = \langle
\theta^2 \rangle /2 \tau_0$. To estimate the
dissipation rate $\varepsilon_\theta$, we apply
the Kolmogorov-Obukhov hypothesis:
$\varepsilon_\theta \approx \langle \theta^2
\rangle / 2 \tau_0$ (see, e.g., Refs.
\cite{MY75,Mc90}), where $\tau_0=\ell/u_0$ is the
characteristic turbulent time and $u_0$ is the
characteristic turbulent velocity at the integral
turbulent scale $\ell$. Indeed,
$\varepsilon_\theta \equiv D \langle
(\bec{\nabla} \theta)^2\rangle = D \, \langle
\theta^2 \rangle \, \int_{k_0}^{k_b} k^2 \,
\tilde E_\theta(k) \,d k \approx \langle \theta^2
\rangle / 2\tau_0$, where $D$ is the coefficient
of molecular temperature diffusion, $\tilde
E_\theta(k)=(2/3) \, k_0^{-1} \, (k/k_0)^{-5/3}$
is the spectrum function of the temperature
fluctuations, $k_0 = \ell^{-1}$, $\, k_b =
\ell_b^{-1}$, $\ell_b=\ell / {\rm Pe}^{3/4}$ and
${\rm Pe} = u_{0} \, \ell / D$ is the Peclet
number. The latter estimate implies that the main
contribution to the dissipation rate
$\varepsilon_\theta$ arises from very small
molecular temperature diffusion scales $\ell_b$.

Substituting the components of the turbulent heat
flux into Eq.~(\ref{AB1}), we obtain the
following equation:
\begin{eqnarray}
{\ell_\ast \, \nabla_\ast \overline{T} \over
\sqrt{\langle \theta^2 \rangle}} = {1 \over 2
C_F} = {\rm const}, \label{A12}
\end{eqnarray}
where
\begin{eqnarray}
[\ell_\ast \, \nabla_\ast \overline{T}]^2 &=&
\big[(\ell_x \nabla_x \overline{T})^2 + (\ell_y
\nabla_y \overline{T})^2 + (\ell_z \nabla_z
\overline{T})^2\big]
\nonumber\\
&& \times \big[1 + 2 C_\theta C_F \beta \tau_0^2
(\nabla_z \overline{T})\big]^{-1} .
 \label{A14}
\end{eqnarray}
In deriving Eq.~(\ref{A12}) we have neglected the
terms $\sim O[\ell^3/L_T^3; \ell^3/(L_T^2 L_U)]$,
where $L_T$ and $L_U$ are the characteristic
spatial scales of the mean temperature and
velocity field variations.

Next, we derive equation for the vertical
turbulent velocity in the stably stratified
turbulent flow. We use the budget equation for
the vertical turbulent kinetic energy
$E_z=\langle u_z^2 \rangle/2$:
\begin{eqnarray}
{DE_z \over Dt} + {\rm div} \,  {\bf \Phi}_{z} =
\langle u_z  f_{z} \rangle +\beta \, F_z + Q_z -
\varepsilon_z ,
 \label{A20}
\end{eqnarray}
(see, e.g., Refs. \cite{KF84,CCH02,ZKR07,ZKR08})
where the third-order moment ${\bf \Phi}_{z} =
\rho^{-1} \langle u_z \, p\rangle \, {\bf e} +
(1/2) \langle {\bf u}  \, u_z^2 \rangle$
determines the flux of $E_z$, $\, \langle u_z
f_{z} \rangle$ is the production term, and $Q_z$
is the inter-component energy exchange term that
according to the ''return-to-isotropy''
hypothesis\cite{R51} is given by
$Q_z=-C_r(E_z-E_k/3)/2\tau_0$. Here we have
neglected a small production term due to the weak
non-uniform mean flow. In the steady-state
Eq.~(\ref{A20}) yields:
\begin{eqnarray}
E_z = {\tau_0 \left[\langle u_z  f_{z} \rangle
+\beta \, F_z \right] \over 1 +
C_r\left(1-1/3A_z\right)},
 \label{A21}
\end{eqnarray}
where $C_r$ is an empirical constant,
$A_z=E_z/E_k$ is the vertical anisotropy
parameter. Using a simple estimate for the
vertical heat flux, $F_z \sim - C \,
\sqrt{\langle u_z^2 \rangle} \, \sqrt{\langle
\theta^2 \rangle}$ (where $C$ is the correlation
coefficient), we arrive at the following equation
for the r.m.s. of the vertical turbulent velocity
in the stably stratified turbulent flow:
\begin{eqnarray}
\sqrt{\langle u_z^2 \rangle} \sim \left[\langle
(u^\ast_z)^2\rangle - C_u \ell_z \, \beta \,
\sqrt{\langle \theta^2 \rangle}\right]^{1/2} \;,
\label{FFA}
\end{eqnarray}
where $C_u$ is an empirical constant to be
determined in the experiment, and we have taken
into account that the characteristic velocity for
the isothermal turbulence
$\langle(u^\ast_z)^2\rangle \sim 2 \tau_0 \langle
u_z f_{z} \rangle /\left[1 +
C_r\left(1-1/3A_z\right)\right]$.

\section{Experimental set-up and instrumentation}

\begin{figure}
\vspace*{1mm} \centering
\includegraphics[width=8cm]{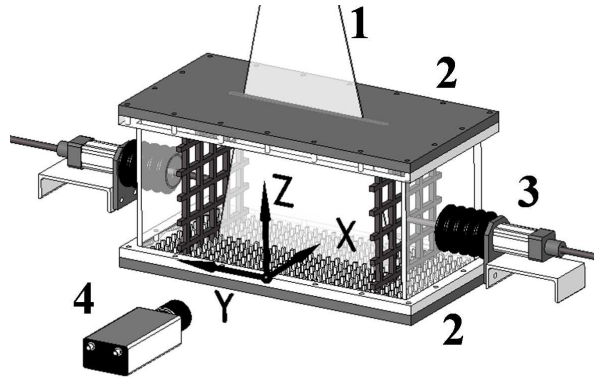}
\caption{\label{Fig1} Experimental set-up: (1)
-laser light sheet in $yz$ plane; (2) - heat
exchangers; (3) - grid driver; (4) - digital CCD
camera.}
\end{figure}

\begin{figure}
\vspace*{4mm} \centering
\includegraphics[width=3cm,angle=-90]{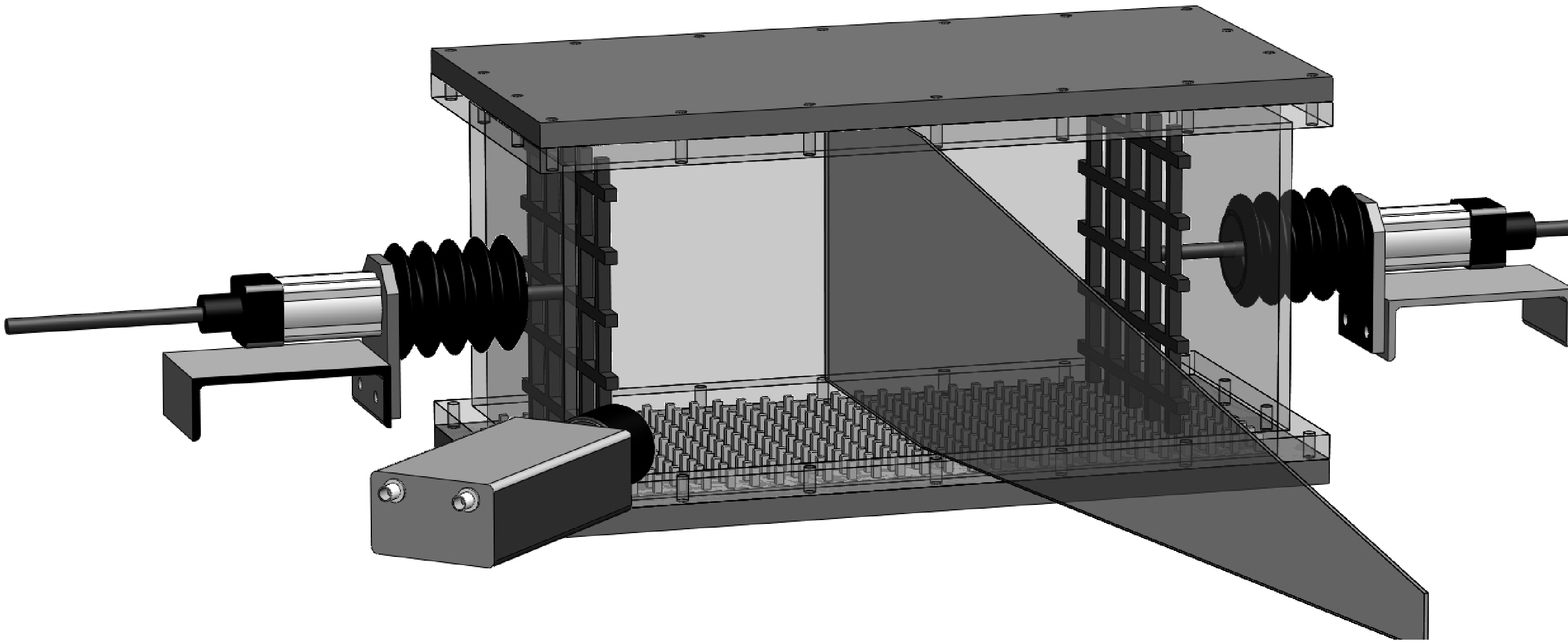}
\caption{\label{Fig2} Experimental set-up with
the laser light sheet in $xz$ plane.}
\end{figure}

\begin{figure}
\vspace*{1mm} \centering
\includegraphics[width=6cm,angle=-90]{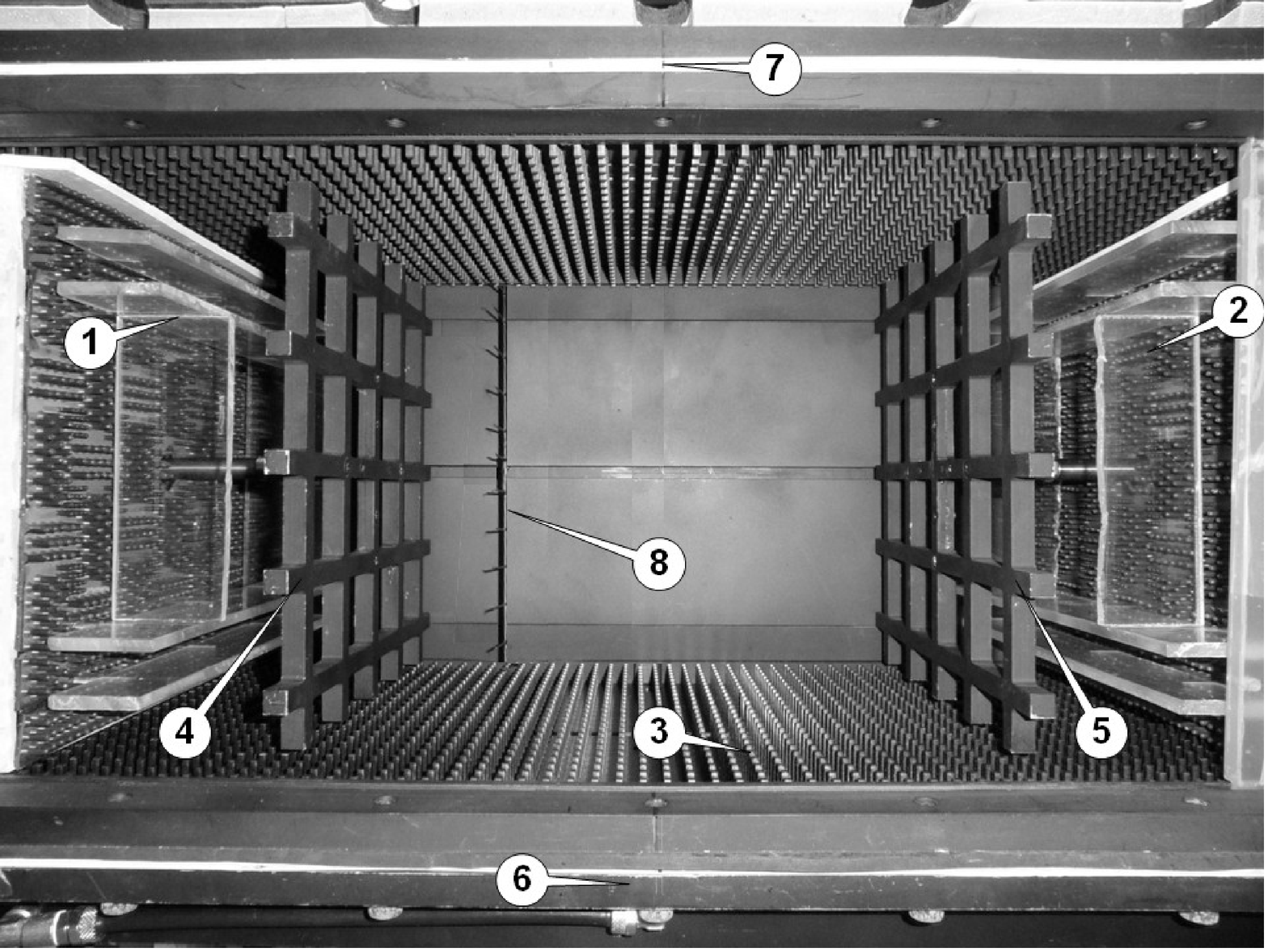}
\caption{\label{Fig3} Rectangular chamber: (1,2)
- partitions ; (3) - rectangular fins; (4,5) -
oscillating grids; (6,7) - heat exchangers; (8) -
temperature measurement system.}
\end{figure}

In this section we describe the experimental
set-up. The experiments in stably stratified
turbulence have been conducted in rectangular
chamber with dimensions $26 \times 58 \times 26$
cm$^3$ in air flow with the Prandtl number Pr
$=0.71$.  The side walls of the chambers are made
of transparent Perspex with the thickness of $1$
cm. In the experiments turbulence is produced by
two oscillating grids. Pairs of vertically
oriented grids with bars arranged in a square
array (with a mesh size 5 cm) are attached to the
right and left horizontal rods (see
Figs.~\ref{Fig1}-\ref{Fig3}). The grids are
positioned at a distance of two grid meshes from
the chamber walls and are parallel to the side
walls. Both grids are operated at the same
amplitude of $3.05$ cm, at a random phase and at
the same frequency which is varied in the range
from $1$ Hz to $10.5$ Hz. To increase the size of
the domain with a homogeneous turbulence and to
decrease the mean velocity, vertical partitions
were attached to the side walls of the chamber.
We use the following system of coordinates: $z$
is the vertical axis, the $y$-axis is
perpendicular to the grids and the $xz$-plane is
parallel to the grids. The aspect ratio of the
chamber $H_y/H_z = 1.52$, where $H_y$ is the size
of the chamber along $y$-axis between partitions
and $H_z$ is the hight of the chamber,
respectively.

A vertical mean temperature gradient in the
turbulent air flow was formed by attaching two
aluminium heat exchangers to the bottom and top
walls of the test section (a cooled bottom and a
heated top wall of the chamber). To improve heat
transfer in the boundary layers at the bottom and
top walls we used heat exchangers with
rectangular fins $0.3 \times 0.3 \times 1.5$
cm$^3$ (see Fig.~\ref{Fig3}) which allowed us to
form a mean temperature gradient up to 1.15 K/cm
at a mean temperature of about 308 K when the
frequency of the grid oscillations $f > 10$ Hz.
The thickness of the aluminium plates with the
fins is 2.5 cm. The bottom plate is a top wall of
the tank with cooling water. Temperature of water
circulating through the tank and the chiller is
kept constant within $0.1$ K. Cold water is
pumped into the cooling system through two inlets
and flows out through two outlets located at the
side walls of the cooling system. The top plate
is attached to the electrical heater that
provides constant and uniform heating. The
voltage applied to the heater varies up to 155 V.
The power of the heater varies up to 300 W.

The temperature field was measured with a
temperature probe equipped with twelve
E-thermocouples (with the diameter of 0.013 cm
and the sensitivity of $\approx 65 \, \mu$V/K)
attached to a vertical rod with a diameter 0.4
cm. The spacing between thermocouples along the
rod was 2.2 cm. Each thermocouple was inserted
into a 0.1 cm diameter and 4.5 cm long case. A
tip of a thermocouple protruded at the length of
1.5 cm out of the case. The temperature in the
central part of the chamber was measured for 2
rod positions in the horizontal and vertical
directions, i.e., at 24 locations in a flow (see
Fig.~\ref{Fig3}). The exact position of each
thermocouple was measured using images captured
with the optical system employed in PIV
measurements. A sequence of 500 temperature
readings for every thermocouple at every rod
position was recorded and processed using the
developed software based on LabView 7.0.

The velocity fields were measured using a
Stereoscopic Particle Image  Velocimetry (PIV),
see Refs. \cite{AD91,RWK07,W00}. In the
experiments we used LaVision Flow Master III
system. A double-pulsed light sheet was provided
by a Nd-YAG laser (Continuum Surelite $ 2 \times
170$ mJ). The light sheet optics includes
spherical and cylindrical Galilei telescopes with
tuneable divergence and adjustable focus length.
We used two progressive-scan 12 bit digital CCD
cameras (with pixel size $6.7 \, \mu$m $\times \,
6.7 \, \mu$m and $1280 \times 1024$ pixels) with
a dual-frame-technique for cross-correlation
processing of captured images. A programmable
Timing Unit (PC interface card) generated
sequences of pulses to control the laser, camera
and data acquisition rate. The software package
LaVision DaVis 7 was applied to control all
hardware components and for 32 bit image
acquisition and visualization. This software
package comprises PIV software for calculating
the flow fields using cross-correlation analysis.

To obtain velocity maps  in the central region of
the flow in the cross-section parallel to the
grids and perpendicular to a front view plane, we
used one camera with a single-axis Scheimpflug
adapter. The angle between the optical axis of
the camera and the front view plane as well as
the angle between the optical axis and the probed
cross-section was approximately 45 degrees (see
Fig.~\ref{Fig2}). The perspective distortion was
compensated using Stereoscopic PIV system
calibration kit whereby the correction was
calculated for a recorded image of a calibration
plate. The corrections were introduced in the
probed cross-section images before their
processing using a cross-correlation technique
for determining velocity fields.

An incense smoke with sub-micron particles
($\rho_p / \rho \sim 10^3)$,  was used as a
tracer for the PIV measurements. Smoke was
produced by high temperature sublimation of solid
incense grains. Analysis of smoke particles using
a microscope (Nikon, Epiphot with an
amplification of 560) and a PM-300 portable laser
particulate analyzer showed that these particles
have an approximately spherical shape and that
their mean diameter is of the order of $0.7
\mu$m. The probability density function of the
particle size measured with the PM300 particulate
analyzer was independent of the location in the
flow for incense particle size of $0.5-1 \,\mu
$m. The maximum tracer particle displacement in
the experiment was of the order of $1/4$ of the
interrogation window. The average displacement of
tracer particles was of the order of $2.5$
pixels. The average accuracy of the velocity
measurements was of the order of $4 \%$ for the
accuracy of the correlation peak detection in the
interrogation window of the order of $0.1$ pixel
(see, e.g., Refs. \cite{AD91,RWK07,W00}).

We determined the mean  and the
r.m.s.~velocities, two-point correlation
functions and an integral scale of turbulence
from the measured velocity fields. Series of 520
pairs of images acquired with a frequency of 1
Hz, were stored for calculating velocity maps and
for ensemble and spatial averaging of turbulence
characteristics. The center of the measurement
region in $yz$ and $xz$-planes coincides with the
center of the chamber. We measured velocity in a
flow domain $25.6 \times 25.6$ cm$^2$ with a
spatial resolution of $1024 \times 1024$ pixels.
This corresponds to a spatial resolution 250
$\mu$m / pixel. The velocity field in the probed
region was analyzed with interrogation windows of
$32 \times 32$ or $16 \times 16$ pixels. In every
interrogation window a velocity vector was
determined from which velocity maps comprising
$32 \times 32$ or $64 \times 64$ vectors were
constructed. The mean and r.m.s. velocities for
every point of a velocity map were calculated by
averaging over 520 independent maps, and then
they were averaged over the central flow region.

The two-point correlation functions  of the
velocity field were determined for every point of
the central part of the velocity map (with $16
\times 16$ vectors) by averaging over 520
independent velocity maps, which yields 16
correlation functions in $x, y$ and $z$
directions. Then the two-point correlation
function was obtained by averaging over the
ensemble of these correlation functions. An
integral scale of turbulence, $\ell$, was
determined from the two-point correlation
functions of the velocity field. In the
experiments we evaluated the variability between
the first and the last 20 velocity maps of the
series of the measured velocity field. Since very
small variability was found, these tests showed
that 520 velocity maps contain enough data to
obtain reliable statistical estimates. We found
also that the measured mean velocity field is
stationary.

The characteristic turbulence time in the
experiments  $\tau_z = 0.2 - 1$ seconds that is
much smaller than the time during which the
velocity fields are measured $(520$ s). The size
of the probed region did not affect our results.
The temperature difference between the top and
bottom plates, $\Delta \overline{T}$, in all
experiments was 50 K.

\section{Experimental results and comparison with the theoretical predictions}

We will start with the discussion of the
experimental results on the temperature
measurements in stably stratified turbulent
flows. To avoid side effects of the grids we
present the experimental results recorded in the
central region of the chamber with the size of
$10 \times 10 \times 10$ cm$^3$. The temperature
was measured at 24 locations in a flow (the
spacing between thermocouples was 2.2 cm). The
separation distance of 2.2 cm between thermal
couples is sufficient to measure the gradients of
the mean temperature. Indeed, the integral scale
of turbulence is about 2 cm (see below), and the
characteristic length scale of the mean
temperature field, $L_T = |\bec{\nabla}
\overline{T} / \overline{T}|^{-1}$, is much
larger than the integral scale of turbulence.

In our study we employ a triple decomposition
whereby the instantaneous temperature $T^{tot}=T
+ \theta$, where $\theta$ are the temperature
fluctuations and $T$ is the temperature
determined by sliding averaging of the
instantaneous temperature field over the time
that is by one order of magnitude larger than the
characteristic turbulence time (for the
temperature difference between the top and bottom
plates $\Delta \overline{T}=50$ K and the
frequency $f=10.5$ Hz of the grid oscillations
the sliding average time is 1.6 s, while the
vertical turbulence time is 0.17 s). This
temperature $T$ is given by a sum, $T
=\overline{T} + \delta T$, where $\delta T$ are
the long-term variations of the temperature $T$
due to the nonlinear temperature oscillations
around the mean value $\overline{T}$. The mean
temperature $\overline{T}$ is obtained by the
additional averaging of the temperature $T$ over
the time 400 s.

In the temperature measurements the acquisition
frequency of the temperature was $1.25$ Hz. The
corresponding acquisition time is $0.8$ s, that
is larger than the characteristic turbulence
time, $0.17$ s, and is much smaller than the
period of non-linear oscillations of the mean
temperature, $12$ s (see below). On the other
hand, the time interval of the one realization of
the temperature field is $400$ s, which
corresponds to 500 data points of the temperature
field over which we perform averaging. Therefore,
the acquisition frequency of temperature is high
enough to provide sufficiently long time series
for statistical estimation of the mean
temperature $\overline{T}$.

\begin{figure}
\vspace*{1mm} \centering
\includegraphics[width=9cm]{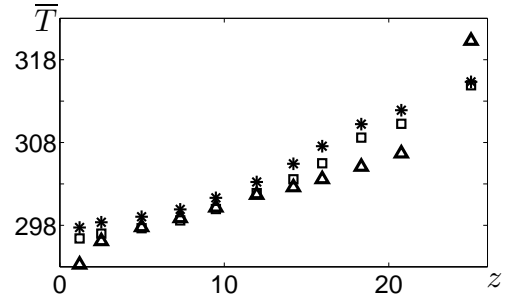}
\caption{\label{Fig4} Vertical profile of the
mean  temperature $\overline{T}(z)$ for different
frequencies $f$ of the grid oscillations for the
stably stratified turbulent flow: $f=1$ Hz
(triangles), $f=2$ Hz (squares), and $f=10.5$ Hz
(snowflakes) for $\triangle \overline{T}=50$ K.
The temperature is measured in K and the distance
in cm.}
\end{figure}

\begin{figure}
\vspace*{1mm} \centering
\includegraphics[width=9cm]{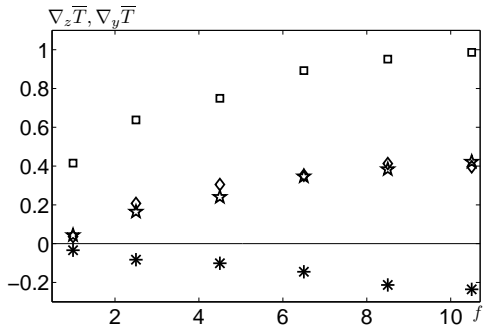}
\caption{\label{Fig5} Vertical gradients of the
mean temperature $\nabla_z \overline{T}$
(squares) and r.m.s. of $\delta (\nabla_z T)$
(diamonds) characterizing amplitude of the
long-term nonlinear oscillations versus the
frequency $f$ of the grid oscillations for the
stably stratified turbulent flow for $\triangle
\overline{T}=50$ K. Similar dependencies are also
shown for horizontal gradients of the temperature
$\nabla_y \overline{T}$ (snowflakes) and r.m.s.
of $\delta (\nabla_y \overline{T})$ (stars). The
temperature gradient is measured in K cm$^{-1}$
and the frequency $f$ is measured in Hz.}
\end{figure}

Let us now analyze the frequency dependence  of
vertical profiles of the mean temperature
$\overline{T}(z)$ (Fig.~\ref{Fig4}). Inspection
of Fig.~\ref{Fig4} shows that the increase of the
frequency $f$ of the grid oscillations weakly
modifies the vertical profiles of the mean
temperature $\overline{T}(z)$. However, the
gradients of the mean temperature $\overline{T}$
in the vertical, $\nabla_z \overline{T}$, and
horizontal, $\nabla_y \overline{T}$, directions
are affected by the increase of the frequency $f$
of the grid oscillations (Fig.~\ref{Fig5}). In
particular, the horizontal and vertical
temperature gradients grow with the frequency of
oscillation. The reasons for that will be
explained in this section.

\begin{figure}
\vspace*{1mm} \centering
\includegraphics[width=9cm]{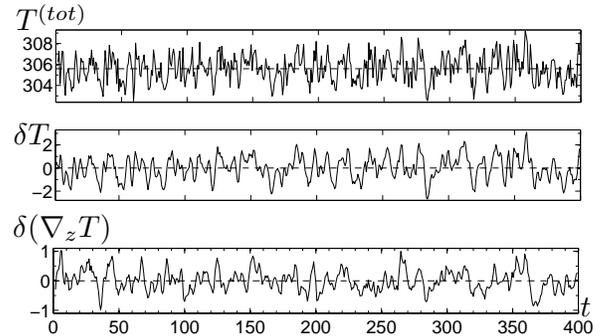}
\caption{\label{Fig6} Time dependencies of the
instantaneous temperature $T^{tot}= T+\theta$,
the variations of mean temperature $\delta T = T
- \overline{T}$ and the variations of the
vertical mean temperature gradient $\delta
(\nabla_z T) = \nabla_z T - \overline{\nabla_z
T}$ due to the long-term nonlinear oscillations
of the mean temperature (with the period $\sim
12$ s). These time dependencies are measured in
the center of the chamber at the frequency
$f=10.5$ Hz of the grid oscillations for the
stably stratified turbulent flow  for $\triangle
\overline{T}=50$ K. These temperature
characteristics are measured in K and time is
measured in seconds.}
\end{figure}

In the experiments we have observed the long-term
nonlinear oscillations $\delta T$ of the
temperature occurring around the mean temperature
$\overline{T}$ with the periods which are much
larger than the turbulent correlation time (see
Fig.~\ref{Fig6}). In particular, in
Fig.~\ref{Fig6} we show time dependencies  of the
instantaneous (actual) temperature $T^{tot}=T +
\theta$, the long-term variations of mean
temperature $\delta T = T - \overline{T}$ and the
long-term variations of the vertical mean
temperature gradient $\delta (\nabla_z T) =
\nabla_z T - \overline{\nabla_z T}$ due to the
nonlinear oscillations of the mean temperature.
We also determined the long-term variations of
the mean temperature gradients $\delta (\nabla_i
T) = \nabla_i T - \overline{\nabla_i T}$ in other
directions, where $i= x, y, z$.

\begin{figure}
\vspace*{1mm} \centering
\includegraphics[width=9cm]{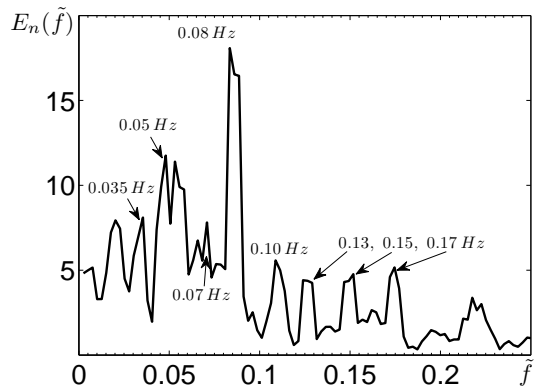}
\caption{\label{Fig7} The normalized spectrum
function $E_n(\tilde f) = |(\delta T)_{\tilde
f}|^2 / \int |(\delta T)_{\tilde f}|^2 \,d\tilde
f$ of the signal $\delta T = T(t) -
\overline{T}$, where in the Fourier space
$(\delta T)_{\tilde f} = \int \delta T \, \exp[-
i  \tilde f \, t] \,d t$ and $\tilde f$ is the
frequency of the nonlinear long-term oscillations
of the mean temperature.}
\end{figure}

In Fig.~\ref{Fig7} we show the results  of a
Fourier analysis of the signal $\delta T = T -
\overline{T}$. Inspection of Fig.~\ref{Fig7}
shows that there are two main maxima in the
spectrum with the periods 12 s and 20 s. Other
smaller maxima in the spectrum are at the
frequencies which are multiples of these main
frequencies or their sums and differences. These
are typical features of nonlinear oscillations.
The theory that explains the mechanism of these
nonlinear oscillations of the mean temperature
field, has not been developed yet. A possible
mechanism for such nonlinear oscillations could
be related to the large-scale
Tollmien-Schlichting waves in sheared turbulent
flows (see Ref. \cite{EGKR07}).

The results of our experiments in the stably
stratified turbulence we will compare with the
results obtained in the similar experimental
set-up, but for the unstably stratified
turbulence or for the isothermal turbulence when
$\triangle \overline{T}=0$. The details of this
experimental set-up and measurements are given in
Ref.\cite{BEKR11}, in which the
Rayleigh-B\'{e}nard apparatus with an additional
source of turbulence produced by the two
oscillating grids located nearby the side walls
of the chamber was used. Additional forcing for
turbulence allows to observe evolution of the
mean temperature and velocity fields during the
transition from turbulent convection with the
large-scale circulations (LSC) for very small
frequencies of the grid oscillations, to the
limiting regime of unstably stratified flow
without LSC for very high frequencies of the grid
oscillations. In the latter case of the unstably
stratified flow without LSC the temperature field
behaves like a passive scalar.\cite{BEKR11}

\begin{figure}
\vspace*{1mm} \centering
\includegraphics[width=9cm]{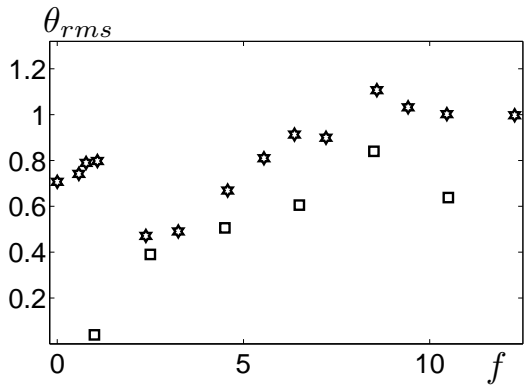}
\caption{\label{Fig8} The r.m.s. of temperature
fluctuations $\theta_{rms}=\sqrt{\langle \theta^2
\rangle}$ versus the frequency $f$ of the grid
oscillations for the stably (squares) and
unstably (stars) stratified turbulent flows.}
\end{figure}

In our experiments with the stably stratified
turbulence, we determined the dependence of the
r.m.s. of the temperature fluctuations
$\sqrt{\langle \theta^2 \rangle}$ versus the
frequency $f$ of the grid oscillations (see
Fig.~\ref{Fig8}), where $\theta$ are fluctuations
of fluid temperature. The temperature
fluctuations monotonically increase with the
increase of the frequency $f$ of the grid
oscillations (except for the higher frequency)
due to the monotonic increase of the mean
temperature gradients. In the case of the
unstably stratified turbulent flow the dependence
$\sqrt{\langle \theta^2 \rangle}$ versus
frequency is more involved.

\begin{figure}
\vspace*{1mm} \centering
\includegraphics[width=10cm]{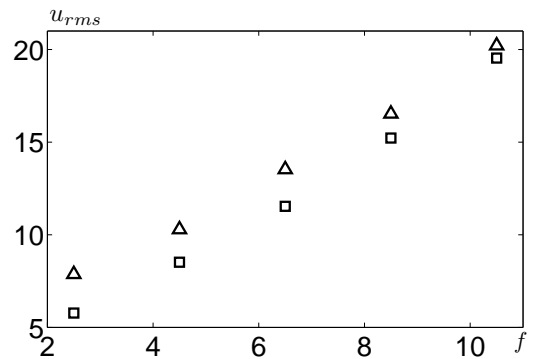}
\caption{\label{Fig9} The r.m.s. turbulent
velocity $u_{rms}$ versus frequency $f$ of the
grid oscillations for the stably stratified
turbulent flow with $\triangle \overline{T}=50$ K
(squares) and for isothermal turbulence
(triangles). The turbulent velocity is measured
in cm s$^{-1}$ and the frequency $f$ is measured
in Hz.}
\end{figure}

\begin{figure}
\vspace*{1mm} \centering
\includegraphics[width=10cm]{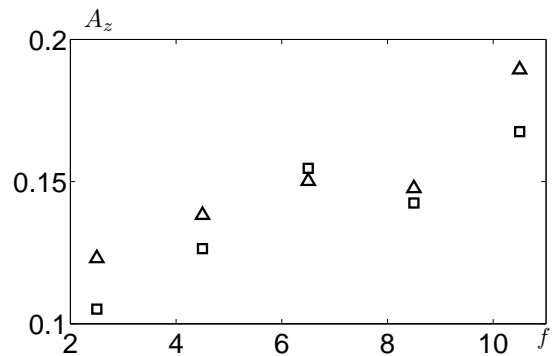}
\caption{\label{Fig10} Vertical anisotropy
$A_z=\langle u_z^2 \rangle/u_{\rm rms}^2$ versus
frequency $f$ of the grid oscillations for the
stably stratified turbulent flow with $\triangle
\overline{T}=50$ K (squares) and for isothermal
turbulence (triangles).}
\end{figure}

\begin{figure}
\vspace*{1mm} \centering
\includegraphics[width=9cm]{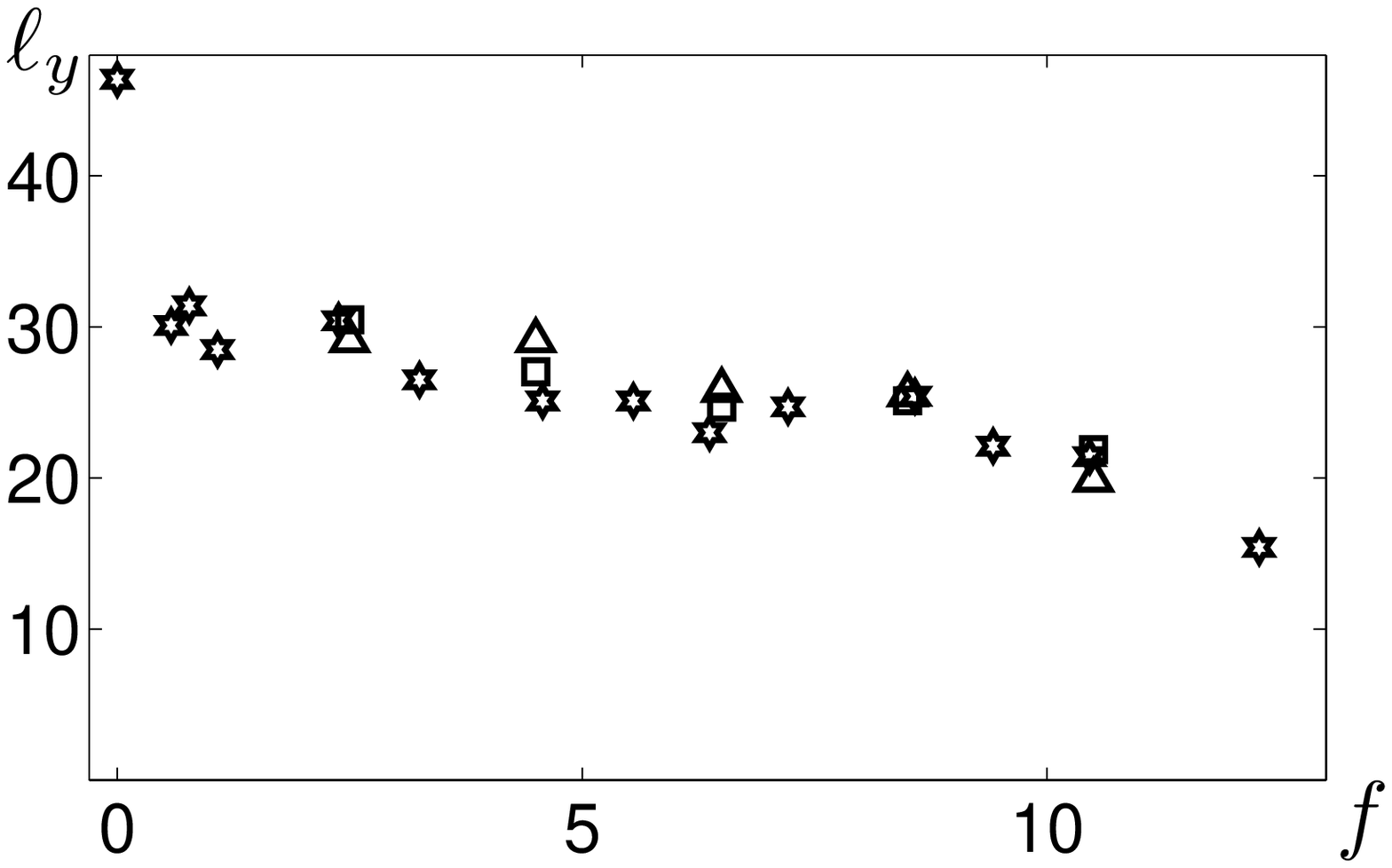}
\caption{\label{Fig11} Horizontal integral scale
of turbulence $\ell_y$ versus the frequency $f$
of the grid oscillations for the stably (squares)
and unstably (stars) stratified turbulent flows
with $\triangle \overline{T}=50$ K, and for
isothermal turbulence (triangles). The turbulent
length scales are measured in mm and the
frequency $f$ is measured in Hz.}
\end{figure}

\begin{figure}
\vspace*{1mm} \centering
\includegraphics[width=9cm]{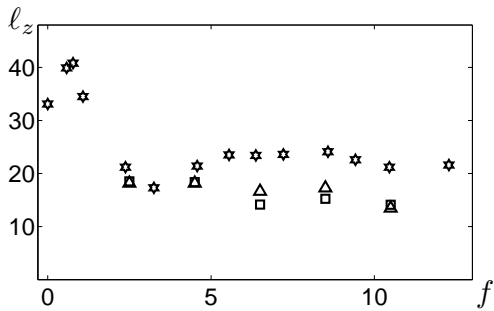}
\caption{\label{Fig12} Vertical integral scale of
turbulence $\ell_z$ versus the frequency $f$ of
the grid oscillations for the stably (squares)
and unstably (stars) stratified turbulent flows
with $\triangle \overline{T}=50$ K, and for
isothermal turbulence (triangles). The turbulent
length scales are measured in mm and the
frequency $f$ is measured in Hz.}
\end{figure}

We also determined the frequency dependencies of
the following measured turbulence parameters: the
r.m.s. velocity fluctuations, $u_{\rm rms}=\left[
\langle u_x^2 + u_y^2 + u_z^2
\rangle\right]^{1/2}$ (Fig.~\ref{Fig9}), the
vertical anisotropy $A_z=\langle u_z^2
\rangle/u_{\rm rms}^2$ (Fig.~\ref{Fig10}), and
the integral scales of turbulence along
horizontal $y$ (Fig.~\ref{Fig11}) and vertical
$z$ (Fig.~\ref{Fig12}) directions $(\ell_y$ and
$\ell_z)$. Except for the small frequencies of
the grid oscillations, the horizontal integral
scale of turbulence behaves similarly for both,
the stably and unstably stratified turbulent
flows, while the vertical integral scale of
turbulence is systematically higher for the
unstably stratified turbulent flow. On the other
hand, the vertical anisotropy $A_z$ only slightly
increases with the frequency of the grid
oscillations.

Note that in the case of unstably stratified
turbulent flow, a cutoff at the frequency of
nearly 1.5 Hz is observed (see Figs.~\ref{Fig8}
and~\ref{Fig11}-\ref{Fig12}). At this frequency
of the grid oscillations the large-scale coherent
structures (convective cells) begin to break down
due to the external forcing.

The two-point correlation functions of the
velocity field have been calculated by averaging
over 520 independent velocity maps (the time
difference between the obtained velocity maps is
by one order of magnitude larger than the
turbulent time scales), and then they have been
averaged over the central flow region. The
integral scales of turbulence, $\ell_y$ and
$\ell_z$, have been determined from the
normalized two-point longitudinal correlation
functions of the velocity field, e.g.,
$F_y(\tilde y)=\langle u_y({\bf r}_0) \, u_y({\bf
r}_0 + \tilde y \, {\bf e}_y)\rangle / \langle
u_y^2({\bf r}_0)\rangle$ [and similarly for
$F_z(\tilde z)$ after replacing in the above
formula $y$ by $z$], using the following
expression: $\ell_y = \langle \int_0^L F_y
(\tilde y) \, d \tilde y \rangle_S$ [and
similarly for $\ell_z$ after replacing in the
above formula $y$ by $z$], where $L=10$ cm is the
linear size of the probed flow region, ${\bf
e}_y$ is the unit vector in the $y$ direction,
and $\langle ... \rangle_S$ is the additional
averaging over the $yz$ cross-section of the
probed region. Since the integral scales of
turbulence, $\ell_y$ and $\ell_z$ are less than 3
cm, the size of the probed region, $L=10$ cm, is
sufficiently large to assure a correct
calculation of the integral scale of turbulence.
We have checked that the increase of the size of
the probed region, does not change the integral
scales of turbulence.

\begin{figure}
\vspace*{1mm} \centering
\includegraphics[width=9cm]{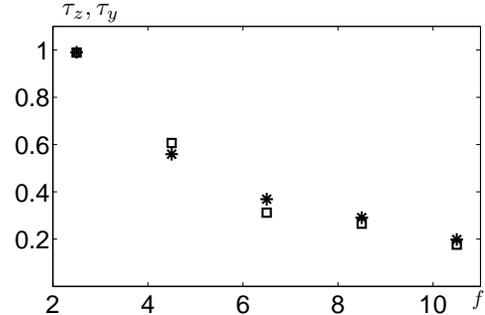}
\caption{\label{Fig13} The turbulent times
$\tau_y=\ell_y/\sqrt{\langle u_y^2 \rangle}$
(snowflakes) and $\tau_z=\ell_z/\sqrt{\langle
u_z^2 \rangle}$ (squares) along horizontal $y$
and vertical $z$ directions versus the frequency
$f$ of the grid oscillations for the stably
stratified turbulent flow with $\triangle
\overline{T}=50$ K. The turbulent times are
measured in s and the frequency $f$ is measured
in Hz.}
\end{figure}

\begin{figure}
\vspace*{1mm} \centering
\includegraphics[width=9cm]{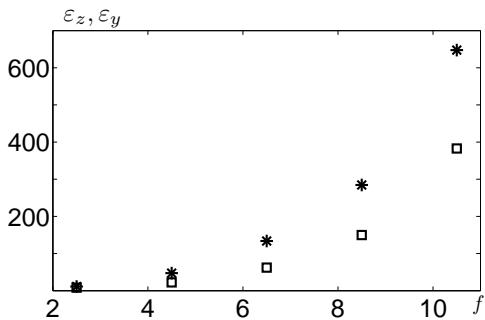}
\caption{\label{Fig14} The rates of dissipation
of the turbulent kinetic energies
$\varepsilon_y=\langle u_y^2 \rangle^{3/2}
/\ell_y$ (snowflakes) and $\varepsilon_z=\langle
u_z^2 \rangle^{3/2} /\ell_z$ (squares) along
horizontal $y$ and vertical $z$ directions versus
the frequency $f$ of the grid oscillations for
the stably stratified turbulent flow with
$\triangle \overline{T}=50$ K. The rates of
dissipation of the turbulent kinetic energies are
measured in cm$^2$ s$^{-3}$ and the frequency $f$
is measured in Hz.}
\end{figure}

In Fig.~\ref{Fig13} we show the turbulent times,
$\tau_y=\ell_y/\sqrt{\langle u_y^2 \rangle}$ and
$\tau_z=\ell_z/\sqrt{\langle u_z^2 \rangle}$,
along horizontal $y$ and vertical $z$ directions
versus the frequency $f$ of the grid
oscillations. When $f > 3$ Hz, the the turbulent
times along horizontal $y$ and vertical $z$
directions nearly coincide. In Fig.~\ref{Fig14}
we also show the rates of dissipation of the
turbulent kinetic energies,
$\varepsilon_y=\langle u_y^2 \rangle^{3/2}
/\ell_y$ and $\varepsilon_z=\langle u_z^2
\rangle^{3/2} /\ell_z$, along horizontal $y$ and
vertical $z$ directions versus the frequency $f$
of the grid oscillations. The difference in the
rates of dissipation of the turbulent kinetic
energies along horizontal $y$ and vertical $z$
directions increases with increase of the
frequency $f$ of the grid oscillations. This is
due to the fact that when  the frequency $f$
increases, the horizontal turbulent velocity
fluctuations increase faster than that in the
vertical direction.

\begin{figure}
\vspace*{1mm} \centering
\includegraphics[width=10cm]{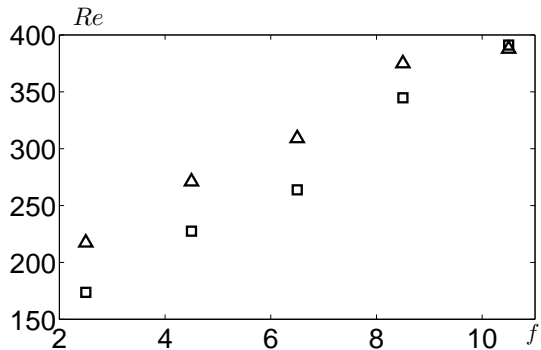}
\caption{\label{Fig15} Reynolds number $Re=\ell
\, u_{rms}/\nu$ versus frequency $f$ of the grid
oscillations for the stably stratified turbulent
flow for $\triangle \overline{T}=50$ K (squares)
and for isothermal turbulence (triangles).}
\end{figure}

In Fig.~\ref{Fig15} we show the Reynolds number
$Re=\ell \, u_{rms}/\nu$ versus frequency of the
grid oscillations for the stably stratified
turbulent flow, where $\ell
=(\ell_x^2+\ell_y^2+\ell_z^2)^{1/2}$. The
Reynolds number increases with the increase of
the frequency of the grid oscillations due to
increase of the production rate of turbulence. On
the other hand, for the largest frequency $f$ the
Reynolds number is independent of the
stratification. This is because for the largest
frequency $f$ of the grid oscillations the
production of turbulence by the grid oscillations
is much larger than suppression of the turbulence
due to the buoyancy. We stress again that the
parameters shown in Figs.~\ref{Fig8}-\ref{Fig15},
have been calculated by the spatial averaging
over the central part of the chamber where the
turbulent flow is nearly uniform.

Now let us explain why the mean temperature
gradients increase with the frequency $f$ of the
grid oscillations (see Fig.~\ref{Fig7}). When the
frequency $f$ of the grid oscillations increases,
the fluctuations of velocity, $u_{\rm rms}$, and
temperature, $\theta_{\rm rms}$, increase, while
the integral scale of turbulence, $\ell$,
decreases (see Figs.~\ref{Fig8}-\ref{Fig9}
and~\ref{Fig11}-\ref{Fig12}). This is the reason
why the turbulent heat flux $(\propto u_{\rm rms}
\, \theta_{ \rm rms})$ increases faster than the
turbulent diffusivity $(D^T \propto u_{\rm rms}
\, \ell)$. Consequently, the mean temperature
gradients $|\nabla_i \overline{T}| \sim |F_i|
/D^T_i$, increase with the frequency $f$ of the
grid oscillations.

\begin{figure}
\vspace*{1mm} \centering
\includegraphics[width=9cm]{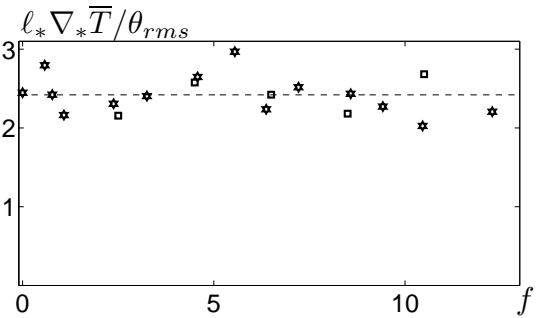}
\caption{\label{Fig16} The non-dimensional ratio
$\ell_\ast \, \nabla_\ast \overline{T} /
\theta_{rms}$ versus the frequency $f$ of the
grid oscillations for the stably (squares) and
unstably (stars) stratified turbulent flows for
$\triangle \overline{T}=50$ K, where
$\theta_{rms}=\sqrt{\langle \theta^2 \rangle}$.}
\end{figure}

In Fig.~\ref{Fig16} we plot the non-dimensional
ratio $\ell_\ast \, \nabla_\ast \overline{T} /
\sqrt{\langle \theta^2 \rangle}$ [see
Eqs.~(\ref{A12}) and~(\ref{A14})] versus the
frequency $f$ of the grid oscillations for the
stably stratified turbulent flow (squares)
obtained in our experiments. For comparison in
the same figure we show also this non-dimensional
ratio obtained previously\cite{BEKR11} for the
unstably stratified turbulent flow (stars).
Inspection of Fig.~\ref{Fig16} shows that this
non-dimensional ratio is nearly independent of
the frequency of the grid oscillations and has
the same magnitude for both, stably and unstably
stratified turbulent flows, in agreement with the
theoretical predictions. Here we assumed that
$\ell_x = \ell_y$, $C_F =0.2$ and $C_\theta =
0.83$ for the unstably stratified turbulence,
while $C_\theta = 1.4$ for the stably stratified
turbulence. Small deviations of the experimental
results from the theoretical predictions [see
Eq.~(\ref{A12})] may be caused by a non-zero
term, ${\rm div} \, {\bf \Phi}_\theta$.

Our measurements showed that $\tau_y =\tau_z$
(see Fig.~\ref{Fig13}), where $\tau_i=\ell_i /
u_i$, and the deviation of the ratio $\tau_z
/\tau_x$ from 1 is small. It should be noted also
that the accuracy of the velocity measurements in
the $x$ direction is probably less than in the
$y, z$ directions, because of the use of
Scheimpflug correction. In the theoretical
estimates for simplicity we assumed that $\tau_x
\approx \tau_y \approx \tau_z =\tau_0$. However,
our main results [Eqs.~(\ref{A12})-(\ref{A14})]
are nearly independent of this assumption since
main contributions to
Eqs.~(\ref{A12})-(\ref{A14}) is from the term
$\propto (\nabla_z \overline{T})^2$.

\begin{figure}
\vspace*{1mm} \centering
\includegraphics[width=9cm]{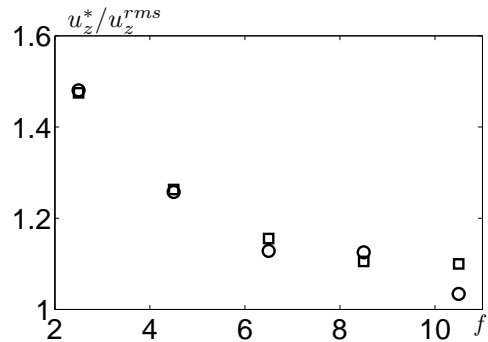}
\caption{\label{Fig17} Ratios
$u^\ast_z/u_z^{rms}$ (squares) and
$u^\ast_z/\tilde u_z$ (circles) versus the
frequency $f$ of the grid oscillations. Here
$u_z^{rms}$ is the r.m.s. of the vertical
component of velocity fluctuations in the stably
stratified turbulent flow, $u^\ast_z$ is the
r.m.s. of the vertical component of velocity
fluctuations in the isothermal turbulence,
$\tilde u_z$ is the vertical component of the
effective turbulent velocity, $\tilde u_z =
[(u^\ast_z)^2 - C_u \ell_z \, \beta \,
\sqrt{\langle \theta^2 \rangle}]^{1/2}$, that
takes into account the attenuation of the
turbulence by buoyancy.  Here $C_u$ is an
empirical constant that is about 1.81 for stably
stratified turbulence and is about 4 for unstably
stratified turbulence. The velocity is measured
in cm s$^{-1}$ and the frequency $f$ is measured
in Hz.}
\end{figure}

We also determined the frequency dependence of
the ratio $u^\ast_z/u_z^{rms}$ (see
Fig.~\ref{Fig17}), where $u_z^{rms}$ is the
r.m.s. of the vertical component of velocity
fluctuations in the stably stratified turbulent
flow and $u^\ast_z$ is the r.m.s. of the vertical
component of velocity fluctuations in the
isothermal turbulence. We also determine the
ratio $u_z^\ast/\tilde u_z$, where $\tilde u_z$
is the vertical component of the effective
turbulent velocity, $\tilde u_z = [(u^\ast_z)^2 -
C_u \ell_z \, \beta \, \sqrt{\langle \theta^2
\rangle}]^{1/2}$. This effective velocity [see
Eq.~(\ref{FFA})] that takes into account the
decay of the turbulence by buoyancy, is derived
from the budget equation~(\ref{A20}) for the
vertical turbulent kinetic energy in Sect.~II.
Inspection of Fig.~\ref{Fig17} shows that the
values of these ratios, $u^\ast_z/u_z^{rms}$ and
$u_z^\ast/\tilde u_z$, are very close. The latter
implies that the measured turbulent velocity in
the stably stratified turbulent flow,
$u_z^{rms}$, is of the order of $\tilde u_z$, in
agreement with the theoretical predictions.

In our experiments the velocity and temperature
fields are not acquired simultaneously. This may
impair the accuracy of the estimates of the
correlation coefficient $C$ in the vertical heat
flux and the empirical constant $C_u$ in
Eq.~(\ref{FFA}). However, in our experiments the
turbulence is stationary in the statistical
sense. Therefore, the estimates of the
correlation coefficient $C$ and the empirical
constant $C_u$ based on our experiments are
reasonable.

\begin{figure}
\vspace*{1mm} \centering
\includegraphics[width=9cm]{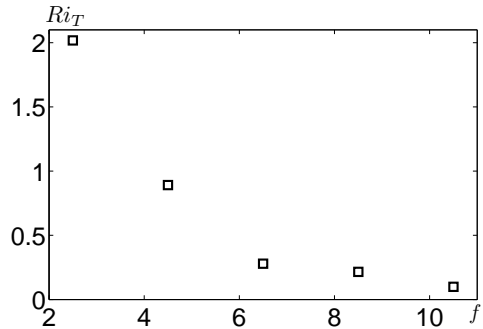}
\caption{\label{Fig18} Turbulent Richardson
number $Ri_T=N^2 \tau_0^2$ versus frequency $f$
of the grid oscillations for the stably
stratified turbulent flow.}
\end{figure}

To characterize the stably stratified flows, in
Fig.~\ref{Fig18} we show the turbulent Richardson
number $Ri_T=N^2 \tau_0^2$ versus frequency $f$
of the grid oscillations for the stably
stratified turbulent flow, where
$\tau_0=\ell/u_{rms}$ is the characteristic
turbulent time. The turbulent Richardson number
$Ri_T$ strongly decreases with the increase of
the frequency of the grid oscillations due to the
strong decrease of the turbulent correlation time
$\tau_0$ with increase of the frequency $f$. For
large frequencies of the grid oscillations
whereby $Ri_T \ll 1$, the temperature field can
be considered as a passive scalar.\cite{YW90} On
the other hand, for smaller frequencies of the
grid oscillations, $Ri_T>1$, and the temperature
field behaves as an active field. Note that the
passive-like scalar behaviour of the temperature
field can be understood in the kinematic sense.
In particular, when the temperature fluctuations
$\langle \theta^2 \rangle$ do not affect the
turbulent kinetic energy, the temperature field
can be considered as a passive scalar. This
implies that the evolution of the temperature
field in a given turbulent velocity field is a
kinematic problem, whereby there is no dynamic
coupling between the temperature fluctuations,
$\langle \theta^2 \rangle$, and the turbulent
kinetic energy. When the effect of the
temperature fluctuations on the turbulent kinetic
energy cannot be neglected, the temperature is
considered as an active field. This definition of
the passive or active behavior of the temperature
field is different from that based on the scaling
behaviour of the temperature structure function.
\cite{LX10}

\section{Conclusions}

Temperature fluctuations in stably stratified
forced turbulence in air flow are investigated in
laboratory experiments. The stratification is
caused by an imposed vertical temperature
gradient, and the turbulence is sustained by
vertical oscillating grids. We demonstrated that
the ratio $\ell_\ast \, \nabla_\ast \overline{T}
/ \sqrt{\langle \theta^2 \rangle}$ determined by
Eq.~(\ref{A14}), is nearly constant and is
independent of the frequency of the grid
oscillations in both, stably and unstably
stratified turbulent flows. We also found that
for large frequencies of the grid oscillations
the turbulent Richardson number, $Ri_T$, is small
and the temperature field can be considered as a
passive scalar, while for smaller frequencies of
the grid oscillations $(Ri_T>1)$ the temperature
field behaves as an active field. The long-term
nonlinear oscillations of the mean temperature in
stably stratified turbulence have been observed
for all frequencies of the grid oscillations
similarly to the case of the unstably stratified
flow. One of the explanations of this effect
could be related to the large-scale
Tollmien-Schlichting waves in sheared turbulent
flows\cite{EGKR07}, which can result in the
nonlinear oscillations of the mean temperature
field.

The temperature fluctuations have been
investigated here also theoretically using the
budget equations for turbulent kinetic energy,
turbulent potential energy (determined by the
temperature fluctuations) and turbulent heat
flux. The developed theory is in a good agreement
with the experimental results.

\begin{acknowledgements}
We thank A.~Krein for his assistance in
construction of the experimental set-up and
J.~Gartner for his assistance in processing of
the experimental results on velocity
measurements. This research was supported in part
by the Israel Science Foundation governed by the
Israeli Academy of Sciences (Grants 259/07 and
1037/11), by EU COST Actions MP0806 and ES1004,
by the EC FP7 project ERC PBL-PMES (Grant 227915)
and the Russian Government Mega Grant implemented
at the University of Nizhny Novgorod (Grant
11.G34.31.0048).
\end{acknowledgements}

\end{document}